\documentstyle[preprint,aps]{revtex}
\begin{document}
\newif\ifplot
\plottrue
\newcommand{\RR}[1]{[#1]}
\newcommand{\intsum}{\sum \kern -15pt \int}
\newfont{\Yfont}{cmti10 scaled 2074}
\newcommand{\Y}{\hbox{{\Yfont y}\phantom.}}
\def\O{{\cal O}}
\newcommand{\bra}[1]{\left< #1 \right| }
\newcommand{\braa}[1]{\left. \left< #1 \right| \right| }
\def\Bra#1#2{{\mbox{\vphantom{$\left< #2 \right|$}}}_{#1}
\kern -2.5pt \left< #2 \right| }
\def\Braa#1#2{{\mbox{\vphantom{$\left< #2 \right|$}}}_{#1}
\kern -2.5pt \left. \left< #2 \right| \right| }
\newcommand{\ket}[1]{\left| #1 \right> }
\newcommand{\kett}[1]{\left| \left| #1 \right> \right.}
\newcommand{\scal}[2]{\left< #1 \left| \mbox{\vphantom{$\left< #1 #2 \right|$}}
\right. #2 \right> }
\def\Scal#1#2#3{{\mbox{\vphantom{$\left<#2#3\right|$}}}_{#1}
{\left< #2 \left| \mbox{\vphantom{$\left<#2#3\right|$}}
\right. #3 \right> }}
\draft
\title{
Scaling properties of the longitudinal and transversal asymmetries of the 
 $\overrightarrow{n}\overrightarrow{d}$ total cross section
}
\author{H. Wita\l a$^*$, W. Gl\"ockle, J. Golak$^*$,  D. H\"uber$^\ddagger$, 
H. Kamada, A. Nogga
}
\address{
$^\dagger$Institut f\"ur theoretische Physik II, Ruhr-Universit\"at Bochum,
D-44780 Bochum, Germany
}
\address{$^{*}$ Institute of Physics, Jagellonian University, 
PL- 30059 Cracow, Poland}
\address{$^\ddagger$Los Alamos National 
Laboratory, M.S. B283, Los Alamos, NM 87545, USA}

\date{\today}
\maketitle
\widetext
\begin{abstract}
The longitudinal and transversal 
asymmetries of the total $\overrightarrow{n}\overrightarrow{d}$ cross section
are calculated. Four modern nucleon-nucleon interactions: 
AV18, CD Bonn, NijmI and 
NijmII, give different predictions for these observables. 
When  the three-nucleon Hamiltonian is supplemented by 
the $2\pi$-exchange Tucson-Melbourne three-nucleon force (3NF), individually 
adjusted  with each particular NN potential to reproduce  
 the  experimental triton binding energy,
  all predictions practically coincide. 
We propose to check this scaling behavior  experimentally in 
order to get a clear signal for 3NF effects in the  low 
energy three-nucleon continuum. 
Connected to that is the proposal to measure the energy at which 
the longitudinal asymmetry goes through zero. This energy is shifted 
by  about 400 keV when 3NF's are acting.
\end{abstract}

\pacs{ PACS numbers: 21.30.+y, 21.45.+v, 24.10.-i, 25.10.+s}
\pagebreak
\narrowtext


Substantial progress in theoretical and numerical methods achieved recently 
in the study of the three-nucleon 
(3N) system allows now to obtain reliable theoretical predictions 
in the 3N continuum which are based on modern nucleon-nucleon (NN) 
interactions, including even three-nucleon forces (3NF)~\cite{8}. 
With the drastically improved quality of the 
present day NN potentials~\cite{2,3,4} 
shown in their 
unprecedented accuracy in reproducing the NN data set 
(${\chi}^2$/datum very close to 1), it is now possible to 
approach the basic and interesting question concerning the 
significance of 3NF's in the elastic nucleon-deuteron (Nd) 
scattering and breakup processes. What is needed are clearcut signals  
coming from certain observables which deny to be explained by 3N 
Hamiltonians based on modern NN interactions  only. The first signature for  
insufficient dynamics based on present day NN forces alone comes from the 3N 
binding energy.    The modern NN interactions underbind $^3H$ by about 
500-800 keV~\cite{6}. In the 3N continuum the bulk of 3N scattering 
observables can be described quite well in the NN force picture only~\cite{8}. 
Recently, however, it was found that a large discrepancy between modern 
NN potential predictions and data exist in the minima of the 
 elastic Nd scattering cross sections for incoming nucleon 
energies greater than $\approx 60$~MeV~\cite{ref2}. A large part of this  
discrepancy can be removed when the $2\pi$-exchange Tucson-Melbourne (TM) 
3NF~\cite{14} properly adjusted to the triton binding is included 
in the 3N Hamiltonian. The question arises if in the  low 
energy 3N-continuum observables exist which also give clearcut signals 
for the action of 3NF's, similar to those coming from 
the $^3He$ and $^3H$ binding energies. The low energy vector analyzing power 
$A_y$ in elastic Nd scattering (as well as $iT_{11}$) could be such a 
case. There a dramatic discrepancies exist between the  
predictions based on NN forces only and both nd and pd 
data~\cite{11},\cite{12}. 
 However, the present day 3NF models have insignificant 
effects and do not remove that discrepancy. This possibly 
implies that it is caused by defects in the 
NN force~\cite{12} or has its source in still undiscovered 3NF 
properties~\cite{hub}. 

In the present study we propose additional candidates for 3NF effects, the 
longitudinal ($\Delta\sigma_L$) and 
transversal ($\Delta\sigma_T$) asymmetries of the total 
neutron-deuteron (nd) cross section with both neutron and deuteron 
polarized. They are defined as 
\begin{eqnarray}
\Delta\sigma_L &=& \sigma_{nd}^{tot}(p_z^n,-p_z^d) - 
\sigma_{nd}^{tot}(p_z^n,p_z^d)   \cr
\Delta\sigma_T &=& \sigma_{nd}^{tot}(p_y^n,-p_y^d) - 
\sigma_{nd}^{tot}(p_y^n,p_y^d),
\label{eq1}
\end{eqnarray}
where  $\sigma_{nd}^{tot}$ is the nd total cross section for  specified 
vector polarizations of the incoming neutron and deuteron
(for instance $p_z^n, (-p_z^d)$ means that the neutron (deuteron) 
is polarized in  z(-z)-direction). The z-axis is 
chosen  parallel to the beam direction and the y-axis  perpendicular 
to the scattering plane. 

Using the optical theorem~\cite{8} it follows that
 the total nd cross section with the incoming polarization state 
described by the density matrix $\rho_{m_n,m_d;m_{n'},m_{d'}}^{in}$ is
given by 
\begin{eqnarray}
\sigma_{tot}^{pol} &=& -{{4m}\over {3}} {1\over {q_0}} (2\pi)^3~
\sum_{m_n m_d m_{n'} m_{d'}}
 Im~[~\rho^{in}_{m_n m_d;m_{n'} m_{d'}} U_{m_{n'} m_{d'};m_n m_d}~], 
\label{eq2a}
\end{eqnarray}
where $U_{m_n^{out},m_d^{out};m_n^{in},m_d^{in}}$ is the forward 
 elastic   nd  
scattering  amplitude at relative  neutron-deuteron momentum $q_0$. 

Inserting the density matrix $\rho^{in}$ for the specific polarization 
states of Eq.(\ref{eq1}) leads to the following expressions for 
the longitudinal and transversal asymmetries
\begin{eqnarray}
\Delta\sigma_L &=& -{{4m}\over {3}} {1\over {q_0}} (2\pi)^3~ p_z^np_z^d~
 Im~[~U_{1/2,-1;1/2,-1} - U_{1/2,1;1/2,1}~]   \cr
\Delta\sigma_T &=&~~ {{4m}\over {3}} {1\over {q_0}} (2\pi)^3 
{1\over {\sqrt 2}}~ 
p_y^np_y^d~ 
 Im~[~U_{-1/2,1;1/2,0} + U_{-1/2,0;1/2,-1}~].
\label{eq2}
\end{eqnarray}

The transition amplitude for elastic Nd scattering is composed of the nucleon 
exchange part ($PG_0^{-1}$), the direct action of a 3NF part 
 $V_4^{(1)}(1+P)$
 and a part 
having its origin in the multiple interactions of 3 nucleons through 
2N and 3N forces:
\begin{eqnarray}
<\phi'\vert U \vert\phi> &=& <\phi'{\vert} PG_0^{-1} +  
 V_4^{(1)}(1+P) + P\tilde T + V_4^{(1)}(1+P)G_0\tilde T{\vert}{\phi}>.
\label{eq10}
\end{eqnarray}
That rescattering part is expressed in terms of a $\tilde T$ operator which 
sums up all multiple scattering contributions through the integral 
equation~\cite{Actap}
\begin{eqnarray}
\tilde T{\vert}{\phi}> &=& tP{\vert}{\phi}> + 
(1+tG_0)V_4^{(1)}(1+P){\vert}{\phi}> 
+ tPG_0\tilde T{\vert}{\phi}> \nonumber \\ 
 &+& (1+tG_0)V_4^{(1)}(1+P)G_0\tilde T{\vert}{\phi}>.
\label{eq20}
\end{eqnarray}
Here $G_0$ is the free 3N propagator, t the NN t-matrix, and P the sum 
of a cyclical and anticyclical permutation of three objects. The 3NF $V_4$ 
is split into 3 parts  
\begin{equation}
V_{4} = \sum_{i=1}^{3} V_4^{(i)}, 
\label{eq3}
\end{equation}
where each one is symmetrical under exchange of two particles. For 
the $\pi-\pi$ exchange 3NF for instance~\cite{14}, this corresponds 
to the three possible choices of the nucleon, which undergoes the (off-shell) 
$\pi-N$ scattering. The asymptotic state ${\vert}{\phi}>$ 
(${\vert}{\phi}'>$) is a product of the deuteron wave function and the 
momentum eigenstate  of the neutron. 

We calculated $\Delta\sigma_L$ and  $\Delta\sigma_T$ at a number of 
neutron energies ranging between 0.6 MeV and 65 MeV by solving Eq.(\ref{eq20}) 
 with four modern NN interactions:  AV18~\cite{3}, CD Bonn~\cite{4}, 
Nijm I and Nijm II~\cite{2}.  As the 3NF we took the $2\pi$-exchange 
TM model~\cite{14} where the strong cut-off 
parameter $\Lambda$ has been adjusted individually together with each 
NN force to the experimental triton binding~\cite{6}. In the 
calculations we included all partial wave states with total 
angular momenta in the two-nucleon subsystem up to $j_{max}=3$. 
For more details on the underlying theoretical formalism and the numerical 
performance we refer to ~\cite{8},\cite{11}. 

We found that the predictions for  $\Delta\sigma_L$ and  $\Delta\sigma_T$ 
obtained with NN interactions only scatter within a range of $\approx 5-10\%$ 
depending on the energy. However, including the 3NF drastically 
reduces  the 
spread of  $\Delta\sigma_L$ and  $\Delta\sigma_T$ values to within a range 
of $\approx 1-2\%$  at incoming 
neutron energies smaller than $\approx 20$MeV. 
This scaling behavior is especially pronounced for $\Delta\sigma_L$, 
where it extents up to $\approx 40$MeV. The scaling behavior 
 diminishes and is finally lost for both 
$\Delta\sigma_L$ and  $\Delta\sigma_T$  
at higher energies. 

Even more interesting is that these scaled values of $\Delta\sigma_L$ 
and  $\Delta\sigma_T$ differ by about $5\%$ from the averaged predictions 
of NN potentials only. This make both asymmetries   good candidates for 
providing a clear 3NF signal detectable by  appropriate   
measurements. In Figs.~1 and 2 we show the ratios of 
$\Delta\sigma_L / p_z^np_z^d$ and $\Delta\sigma_T / p_y^np_y^d$ 
to their average value (averaged over four potential results  
 when 2N and  3NF's are acting). 
Both cases are shown: pure NN force predictions and NN + 3NF predictions. 
 In Fig.~1 it is
 seen  that for $\Delta\sigma_L$ 
 the influence of the 3NF is especially large around $\approx 12$~MeV where 
$\Delta\sigma_L$ changes its sign. 
 At 12 MeV the pure 2N force predictions for 
 ${\Delta}{\sigma}_L^{2NF}/<{\Delta}{\sigma}_L^{2NF+3NF}>$ 
(not shown in Fig.~1), are negative; (the corresponding values are 
-0.84, -0.48, -0.87, and -0.84 for AV18, CDBonn, NijmI and NijmII, 
respectively) while adding the 3NF  positive predictions result. 
However, even in  this energy region 
the scaling behaviour is not destroyed. 
At 12 MeV only the AV18+3NF prediction deviates by about 15\% from a 
practically 
overlapping results for the other three potentials. This deviation, 
however, is 
small compared to the changes caused by adding the 3NF 
to the pure 2N force predictions at this energy.

Another interesting point is that the pure 2N force predictions 
lead to the energy
 $E_n^{cross}(2N)\approx 11.8$~MeV at which ${\Delta}{\sigma}_L$ changes 
sign. This value is different by 
about 400 keV when the 
3NF is acting. Then ${\Delta}{\sigma}_L$ 
crosses zero at   $E_n^{cross}(2N+3NF)\approx 12.2$~MeV. The magnitude of 
that energy shift is  large enough 
to be detected by the measurement of 
 $\Delta\sigma_L$.

The situation is similar for ${\Delta}{\sigma}_T$ as shown in   Fig.~2. 
There is, however, no zero crossing.

Summarizing, we have shown that the 3NF changes the magnitudes 
of the longitudinal and transversal 
asymmetries of the total $\overrightarrow{n}\overrightarrow{d}$
 cross sections. These changes show at low energies scaling 
character bringing different predictions based on modern 2N potentials 
together when the proper 3NF is added, which together  with the particular 
NN interaction reproduces the experimental triton binding energy. 
The magnitude of the effect is of the order of 5\%-10\%, which is large 
enough to be measured. In addition, the zero-crossing energy of 
the longitudinal asymmetry is shifted by  about 400 keV when the 
3NF is acting.  
This  shift is also sufficiently large to be checked experimentally. 
The results of such measurements would form a data basis to test the 
present day 3NF models in the low energy region of the 3N continuum.

\begin{center}
{\bf Acknowledgements}
\end{center}

This work was supported by the Deutsche Forschungsgemeinschaft 
 under Project No. Gl87/24-1, and by the U.S. Department of Energy, 
Office of High Energy and Nuclear Physics under grant No. 
DE-FG02-97ER41033. 
 The numerical calculations have been performed on the 
CRAY T90 and the CRAY T3E of the H\"ochstleistungsrechenzentrum in J\"ulich,
Germany, on the Cray T916 of the North Carolina Supercomputing Center 
at the Research Triangle Park, North Carolina, 
and on the Convex 3820 of the ACK in Cracow, Poland 
(KBN/SPP/UJ/046/1996). We would like especially to thank 
Professor W. Tornow from Duke University for bringing this problem to our 
attention  and for his strong support during 
the performance of this work.

\newpage
\subsection*{Figure Captions}
\begin{description}

\item{Fig.1}
{The ratios of  $\Delta\sigma_L$ to their average 
 value (see text)  as a function of incoming neutron lab. energy. 
The open circles, open squares, open triangles, and crosses 
are the AV18, CDBonn, 
NijmI, and NijmII potential predictions, respectively. 
The full circles connected with a long dashed line, 
the full squares connected with a solid 
line, the full triangles connected with a dashed line, and 
the pluses connected with a short dashed line are the corresponding 
predictions when the 3NF is included.
}

\item{Fig.2}
{The same as in Fig.1 but for   $\Delta\sigma_T$.
}

\end {description}

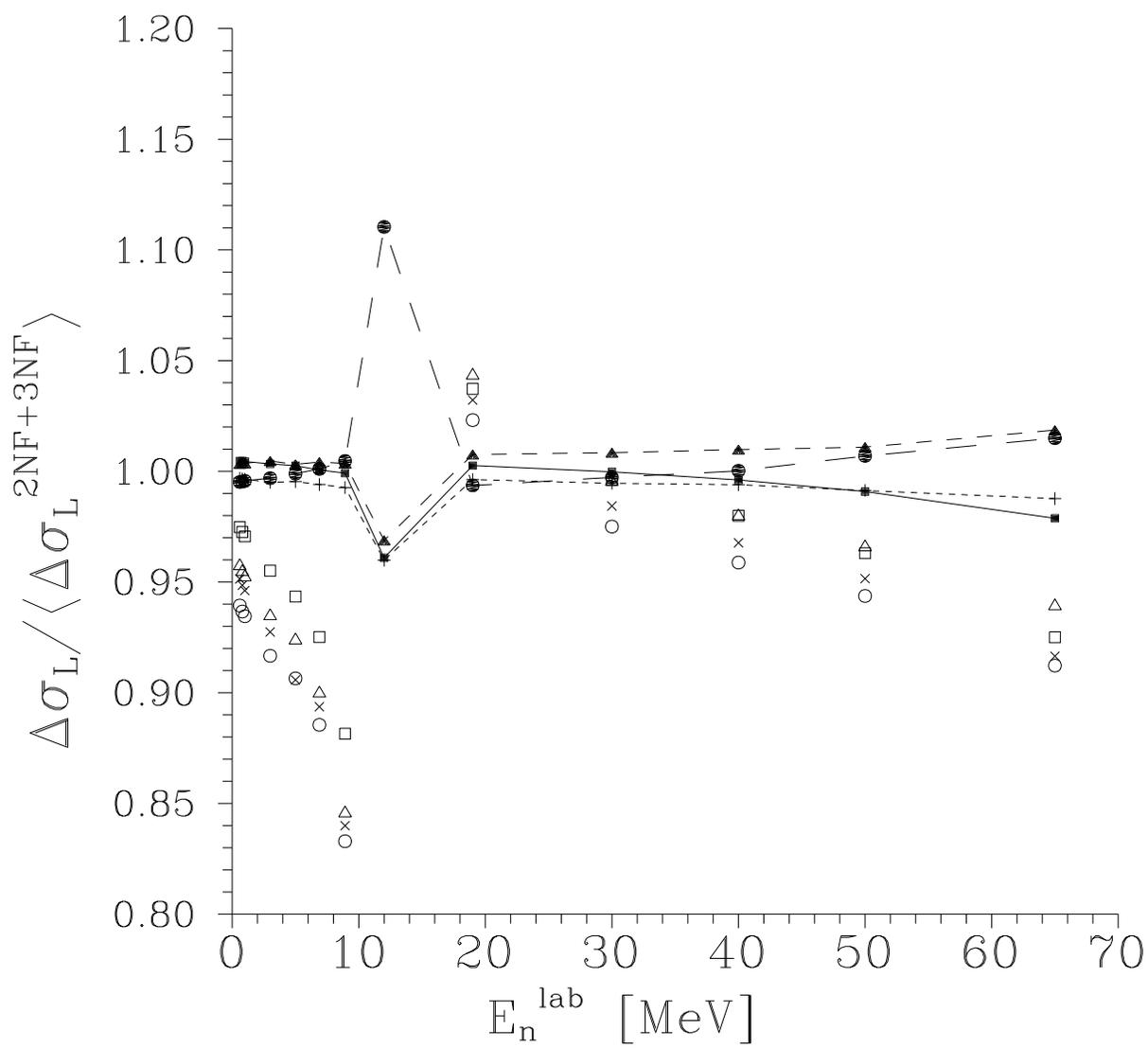
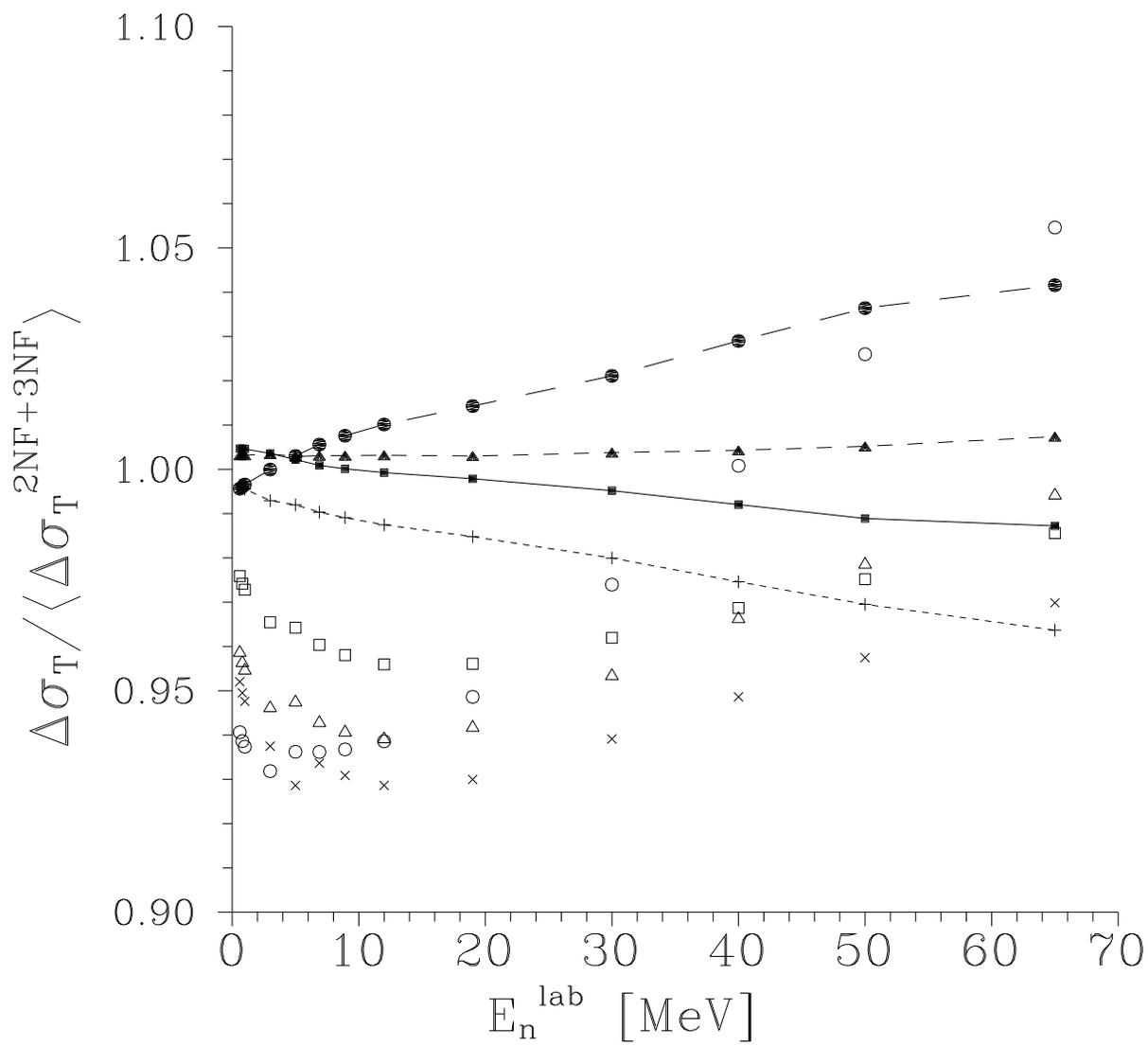
\end{document}